\documentclass[%
 aip,
 amsmath,amssymb,
 reprint,%
]{revtex4-2}
\usepackage{hyperref}
\hypersetup{colorlinks=true, citecolor=blue, urlcolor=blue, linkcolor=blue}

\usepackage{graphicx}
\usepackage{dcolumn}
\usepackage{bm}
\usepackage{xcolor}
\usepackage[normalem]{ulem}
\usepackage{physics}
\usepackage{chemformula}
\let\ket=\pket 
\let\bra=\pbra 
\usepackage{braket}
\usepackage{tikz}
\usetikzlibrary{external}
\usepackage{float}

\DeclareMathOperator*{\argmin}{arg\,min}
\begin{document}

\preprint{}

\title{Accelerated coupled cluster calculations with Procrustes orbital interpolation}

\author{Simon Elias Schrader}
 \email{s.e.schrader@kjemi.uio.no}
\author{Simen Kvaal}%
 \email{simen.kvaal@kjemi.uio.no}

\affiliation{Hylleraas Centre for Quantum Molecular Sciences, Department of Chemistry, University of Oslo, P.O. Box 1033 Blindern, N-0315 Oslo, Norway 
}%

\date{\today}

\begin{abstract}\small
The coupled cluster method is considered a gold standard in quantum chemistry, reliably giving energies that are exact within chemical accuracy (1.6 mHartree). However, even in the CCSD approximation, where the cluster operator is truncated to include only single and double excitations, the method scales as $O(N^6)$ in the number of electrons, and the cluster operator needs to be solved for iteratively, increasing computation time. Inspired by eigenvector continuation, we present here an algorithm making use of Gaussian processes that provides an improved initial guess for the coupled cluster amplitudes. The cluster operator is written as a linear combination of sample cluster operators which are obtained at particular sample geometries. By reusing the cluster operators from previous calculations in that way, it is possible to obtain a start guess for the amplitudes that surpasses both MP2-guesses and "previous geometry"-guesses in terms of the number of necessary iterations. As this improved guess is very close to the exact cluster operator, it can be used directly to calculate the CCSD energy to chemical accuracy, giving approximate CCSD energies scaling as $O(N^5)$.
\end{abstract}

\maketitle

\section{Introduction}\label{sec:Introduction}

{C}oupled cluster (CC) theory \cite{bible,CC_Review_Bartlett} is one of the most used methods in quantum chemistry for the accurate prediction of the structure and the properties of molecular systems. It represents an excellent compromise between accuracy and computational cost, scaling as $O(N^6)$ in the CCSD approximation, with $N$ representing the number of electrons. While calculating the energy of a molecule with CCSD only scales as $O(N^4)$, obtaining the cluster operator scales as $O(N^6)$ \cite{stanton1991direct}. 
Furthermore, the cluster operator is usually obtained iteratively, leading to a total scaling of $O(kN^6)$ with $k$ being the number of iterations. Reducing the number of necessary iterations by having an improved start guess, or having a start guess that can approximate the CCSD energy within chemical accuracy to the converged CCSD energy, can thereby lead to a large speed up in the calculation of coupled cluster energies. Inspired by eigenvector continuation \cite{EVC_math,Eigenvectorcontinuation,EkstromHagen}, where the target wave function is written as a linear combination of sample wave functions, we present here two algorithms which approximate the cluster operator at molecular target \mbox{geometry} $\vec{R}_\odot$ as a linear combination of cluster operators at $M$ sample geometries $\vec{R}_m$, $m=1,\dots,M$, where $\vec{R}$ stands for the position of all nuclei. We call this procedure amplitude coupled cluster eigenvector continuation (\emph{AMP-CCEVC}). The first algorithm makes use of machine learning in the form of Gaussian processes, while the second algorithm corresponds to a simplification of the direct product decomposition by Stanton et al. \cite{stanton1991direct}. {This article shares some parallels with the data-driven prediction of the cluster amplitudes by Townsend and Vogiatzis \cite{Townsend2019}, which also uses machine learning to predict an improved start guess for the cluster operator. That approach predicts each cluster amplitude individually, based on features such as the MP2-amplitudes and orbital energies, whereas our approach predicts the cluster operator as a whole, as a linear combination of sample cluster operators.} While this article focuses on molecular geometries, AMP-CCEVC is not restricted to molecular geometries. It can readily be used for any parameter $\boldsymbol{x}$ that a many-body Hamiltonian $\hat{H}(\mathbf{x})$ is a continuous parameter of.\\
This article is structured as follows: In section \ref{sec:Methods}, a short recap of the CCSD method is given, focusing particularly on the choice of the starting guess for the calculation of the CC amplitudes. After that, we describe the AMP-CCEVC method, where the cluster operator is written as a linear combination of sample cluster operators. We describe the two methods how this linear combination is obtained and give a short overview over Gaussian processes there. This is followed by a very short discussion on orbital connections, and how wave functions at different molecular geometries can be considered to be part of the same Fock space. Then, we consider a choice of orbitals that is suitable for AMP-CCEVC, which we termed \emph{Procrustes orbitals}. Results are presented in section \ref{sec:Results}, where some examples on how AMP-CCSD performs in terms of the number of necessary iterations and energy, are presented. As test molecules, we consider two one-dimensional problems, the dissociation of \ch{HF} in the cc-pVTZ basis set \cite{cc-basis_sets} and the dissociation of the ethylene double bond in the cc-pVDZ basis set \cite{cc-basis_sets}, as well as one two-dimensional problem, the asymmetric stretch of the linear \ch{BeH2} molecule in the cc-pVTZ basis set. This is followed by a discussion in section \ref{sec:Discussion}, where we consider how AMP-CCEVC can be used, and the situations in which it cannot be applied. Our findings are summarized in the conclusion. \\
AMP-CCEVC and Procrustes orbitals have been developed by one of the authors (Simon Elias Schrader) as a part of his master's thesis \cite{SimonMaster}. 

\section{Methods}\label{sec:Methods}
\subsection{CCSD}
In coupled cluster (CC) theory, the wave function takes the form
\begin{equation}
    \ket{\Psi}= e^{\hat{T}}\ket{\Phi_0}=\sum_{n=0}^\infty \frac{1}{n!}\hat{T}^n \ket{\Phi_0}.
\end{equation}
The \emph{cluster operator} $\hat{T}$ is an excitation operator of the form
\begin{equation}
    \hat{T}=\sum_{i=1}^\infty T_i=\sum_{\mu}t_\mu \hat {\tau}_\mu =\sum_{ia}t_i^aa_a^\dagger a_i+\sum_{iajb}\frac{1}{4}t_{ij}^{ab}a_a^\dagger a_b^\dagger a_ja_i+\dots
\end{equation}
acting on the reference determinant $\ket{\Phi_0}$. $\hat \tau_\mu\ket{\Phi_0}=\ket{\mu}$ represents an excited determinant of any order, and we follow Helgaker's notation \cite{bible}, with $i,j,k\dots$ standing for occupied orbitals, $a,b,c,\dots$ standing for virtual orbitals, and $p,q,r,\dots$ for either. In the CCSD model, the cluster operator is truncated to include only single and double excitations.\\ We denote the collection of all amplitudes $\{t_i^a,t_{ij}^{ab},\dots\}$ by a vector $\mathbf{t}$. By considering the operator $\bar{H}=e^{-\hat{T}}\hat{H}e^{\hat T}$, the so-called \emph{similarity-transformed Hamiltonian}, it is straightforward to show, if $\ket{\Psi}=e^{\hat T}\ket{\Phi_0}$ were a true eigenfunction of the Hamiltonian,  that the \emph{energy equation}
\begin{equation}\label{eq:CCenergy}
     \bra{\Phi_0}\bar H\ket{\Phi_0}=E
\end{equation}
and the \emph{amplitude equations}
\begin{equation}\label{eq:CCamplitudes}
     \bra{\mu}\bar H\ket{\Phi_0}=0 \text{ for  $\mu\in S\cup D\cup\dots$}.
\end{equation}
would be satisfied, where $S$ is the set of all singly excited determinants, $D$ the set of all doubly excited determinants, etc.
The cluster amplitudes are determined by solving the amplitude equations \eqref{eq:CCamplitudes}. For CCSD, only singly and doubly excited determinants are considered. As there is no closed-form solution for the amplitudes, one usually resorts to iterative methods, such as quasi-Newton methods, which require an initial guess to be provided. Using canonical orbitals, it is common to use the amplitudes from the first order correction to the reference determinant  $\mathbf{t}^{(1)}$, which we will refer to as the \emph{MP2 guess}:
\begin{equation}
\begin{split}
    {t_i^a}^{(1)}&=0\\
    {t_{ij}^{ab}}^{(1)}&=\frac{\braket{ij||ab}}{\epsilon_a+\epsilon_b-\epsilon_i-\epsilon_j}
\end{split}
\end{equation}
where $\epsilon_p$ is the orbital energy of orbital $p$. Only when the first-order correction to the reference determinant is a good approximation to the true wave function, $\mathbf{t}^{(1)}$ is a sensible initial guess, otherwise, convergence can be slow or even fail. An alternative initial guess is to use amplitudes from a "related" calculation, e.g. a cluster operator at a nearby molecular geometry or a similar field strength. \\ 
In CC theory, the reference determinant $\ket{\Phi_0}$ is usually chosen as the Hartree--Fock ground state and canonical orbitals are used for both the occupied and the virtual orbitals. Similarly to the unitary freedom for Slater determinants, which is invariant to orbital rotations between occupied orbitals, the CC wave function is invariant to rotations happening within the set of occupied orbitals and the set of virtual orbitals, respectively (but not between occupied and virtual orbitals), such that the same wave function can be obtained with other choices of orbitals (leading, though, to a different amplitude vector $\mathbf{t}$). By Thouless' theorem \cite{thouless}, CC theory is relatively insensitive to the choice of the reference determinant $\ket{\Phi_0}$ \cite{Shavitt_Bartlett}, even when the reference state is different from a Hartree--Fock ground state. However, when there are large differences in two possible reference Slater determinants $\ket{\Phi_0}$ and $\ket{\tilde \Phi_0}$, the obtained CCSD wave function and energy will still be qualitatively different. 
\subsection{AMP-CCEVC}
Let the Hamiltonian $\hat{H}(\mathbf{x})$ be dependent on {a (multivariate) parameter $\mathbf{x}\in \mathbb{R}^M$} for some integer $M\geq 1$. A particular example is nuclear positions $\vec{R}\in \mathbb{R}^{3N_A}$ in the Born-Oppenheimer approximation, where $N_A$ is the number of nuclei. Define a set of distinct sample parameters $\{\mathbf{x}_m\}_{m=1}^L$ with corresponding CC wave functions $e^{\hat{T}_m}\ket{\Phi_0}=e^{\hat{T}(\mathbf{x}_m)}\ket{\Phi_0}$. The \emph{AMP-CCEVC} wave function $\ket{\tilde\Psi(\mathbf{x})}$ is then parameterized as
\begin{equation}\label{eq:ClusterOperator}
    \ket{\tilde\Psi(\mathbf{x})}=e^{\tilde{T}(\mathbf{x})}\ket{\Phi_0}=\exp{\sum_{m=1}^Lc_m(\mathbf{x})\hat{T}(\mathbf{x}_m)}\ket{\Phi_0},
\end{equation}
e.g. the true cluster operator $\hat{T}(\mathbf{x})$ is approximated as a linear combination of sample cluster operators $\hat{T}(\mathbf{x}_m)$, and the problem is reduced to finding the coefficients $c_m(\mathbf{x})$ for $m=1,\dots,L$. {We remark that} the amplitude vectors $\mathbf{t}_1,\dots,\mathbf{t}_L$ should be orthonormalized. {This orthonormalization is not necessary for the derivation of any of the theoretical results, but it is necessary for computational reasons -- if two sample amplitude vectors are very similar, $\mathbf{t}_i\approx \mathbf{t}_j$, $i\neq j$, which might be the case when two sample parameters $\mathbf{x}_i$, $\mathbf{x}_j$ are very close, the parameters $c_i(\mathbf{x}),c_j(\mathbf{x})$ can become very large or vary strongly in some region. For example, at the sample parameters $\{\mathbf{x}_i\}_{m=1}^L$, we have that $c_i(\mathbf{x}_j)=\delta_{ij}$. Thus, as $\mathbf{x}_i\rightarrow \mathbf{x}_j$, $i\neq j$, $c_i(\mathbf{x})$ becomes very steep between $\mathbf{x}_i$ and $\mathbf{x}_j$, thereby becoming less well-behaved. Orthonormalization "smoothens" the resulting functions $\{c_m(\mathbf{x})\}_{m=1}^L$, thus improving the methods described below. For the orthonormalization procedure, we have found that using Löwdin symmetric orthogonalization works well, as it, unlike Gram-Schmidt orthonormalization, guarantees that all sample cluster operators have relevant contributions for different values of $\mathbf{x}$.} We will now discuss two algorithms to calculate the set of coefficients $\{c_m(\mathbf{x})\}^L_{m=1}$.
\subsubsection{Learning with Gaussian processes}
Post-orthonormalization, the coefficients $c_n(\mathbf{x}_m)$ for $n=1,\dots,L$, $m=1,\dots,L$ are available. We can consider the $L$ functions $c_n(\mathbf{x})$ as smooth functions to be determined, with observations $c_n(\mathbf{x}_m)$. This is a regression problem which can be tackled with machine learning algorithms. We label the estimator for the true function $c_n(\mathbf{x})$ as $\hat{c}_n(\mathbf{x})$. Gaussian processes (GPs) return the correct values at the sample points: $\hat{c}_n(\mathbf{x}_m)=c_n(\mathbf{x}_m)$, which is a desirable feature, and are known to be suitable for modelling functions with the underlying shape not known. They also provide uncertainty estimates, which makes it possible to design a greedy algorithm that provides an estimate of the "best" sample parameter when more sample parameters are required. In the remainder of this paragraph, we provide a short introduction to what Gaussian processes are and how they are used to estimate parameters. For more information about Gaussian processes, see, for example, any of Refs. \cite{williams2006gaussian,deringer2021gaussian,Schulz_GP,Wang_GP}.
In regression, the aim is to obtain a function
${f(\mathbf{x})}$ that best matches a set of $m$ observations $\{\mathbf{x}_i,f(\mathbf{x}_i)\}_{i=1}^m$ (assuming no noise). Gaussian process regression is a non-parametric regression method that models the probability distribution over all functions ${f(\mathbf{x})}$ that fit the observations. That is, $f(\mathbf{x)}$ is assumed to be distributed as a Gaussian process. A Gaussian process has, by definition, the property that any finite number of random variables follow a multivariate normal distribution. It is a distribution over \emph{functions} 
\begin{equation}
    f(\mathbf{x})\sim \mathcal{GP}(m(\mathbf{x}),k(\mathbf{x},\mathbf{x}'))
\end{equation}
with mean $m(\mathbf{x})$ and covariance $k(\mathbf{x},\mathbf{x}')$. $m(\mathbf{x})$ is the expected value at input $\mathbf{x}$, and $k(\mathbf{x},\mathbf{x}')$ expresses how the output at different input parameters is related:
\begin{equation}
    k(\mathbf{x},\mathbf{x}')=\mathbb{E}(f(\mathbf{x})f(\mathbf{x}'))-m(\mathbf{x})m(\mathbf{x}')
\end{equation}
For computational convenience, one assumes $m(\mathbf{x})=0$ (which can be achieved by subtracting the observed mean), and only a covariance function (usually referred to as \emph{kernel}) needs to be specified. In this article, we use the popular Radial Basis Function (RBF) kernel \begin{equation}\label{RBFkernel}
    k(\mathbf{x},\mathbf{x}')=\sigma_f^2\exp(-\frac{d(\mathbf{x},\mathbf{x}')^2}{2l^2})
\end{equation}
where $\sigma_f, l$ are real hyper-parameters to be specified and 
\begin{equation}\label{eq:distance}
    d(\mathbf{x},\mathbf{x}')=\norm{h(\mathbf{x})-h(\mathbf{x}')}
\end{equation}
for some norm $\norm{\cdot}$ and some function $h(\cdot)$. The fact that, for Gaussian processes, a finite number of random variables have a multivariate normal distribution is used to incorporate observations and make predictions. Denoting two sets of parameters as $\{\mathbf{x}_{i}\}_{i=1}^n$ and $\{\mathbf{x}_{i}'\}_{i=1}^m$ and collecting them column-wise in matrices $\mathbf{X}$ (with dimension $n\times p$) and $\mathbf{X}'$ (with dimension $m\times p$), we define the $n \times m$ covariance matrix
\begin{equation}
    K\left(\mathbf{X}, \mathbf{X}'\right)\!=\!\!\left[\begin{array}{cccc}
\!\!k\left(\mathbf{x}_{1}, \mathbf{x}_{1}'\right) \!& \!k\left(\mathbf{x}_{1}, \mathbf{x}_{2}'\right) & \ldots \!& \!k\left(\mathbf{x}_{1}, \mathbf{x}_{m}'\right)\!\! \\
\!\!k\left(\mathbf{x}_{2}, \mathbf{x}_{1}'\right) \!& \!k\left(\mathbf{x}_{2}, \mathbf{x}_{2}'\right) & \ldots \!& \!k\left(\mathbf{x}_{2}, \mathbf{x}_{m}'\right)\!\! \\
\!\!\vdots & \vdots & \ddots & \vdots\!\! \\
\!\!k\left(\mathbf{x}_{n}, \mathbf{x}_{1}'\right) \!& \!k\left(\mathbf{x}_{n}, \mathbf{x}_{2}'\right) & \ldots \!& \!k\left(\mathbf{x}_{n}, \mathbf{x}_{m}'\right)\!\!
\end{array}\right].
\end{equation}
Denoting the training input $\mathbf{X}_t$ with training output $\mathbf{y}_t$ and the new input as $\mathbf{X}_\odot$, we are interested in the distribution of the targets  $\mathbf{f}_\odot$. They follow, by definition, a multivariate normal distribution 
\begin{equation}
    \left[\begin{array}{l}
\mathbf{y}_t \\
\mathbf{f}_{\odot}
\end{array}\right] \sim \mathcal{N}\left(\mathbf{0},\left[\begin{array}{cc}
K\left(\mathbf{X}_t, \mathbf{X}_t\right) & K\left(\mathbf{X}_t, \mathbf{X}_{\odot}\right) \\
K\left(\mathbf{X}_{\odot}, \mathbf{X}_t\right) & K\left(\mathbf{X}_{\odot}, \mathbf{X}_{\odot}\right)
\end{array}\right]\right)
\end{equation}
where the notation $\mathbf{y}\sim \mathcal{N}\left(\boldsymbol{\mu},\boldsymbol{\Sigma} \right)$ means that the multivariate random vector $\mathbf{y}$ is normally distributed with mean $\boldsymbol{\mu}$ and covariance matrix $\boldsymbol{\Sigma}$.
The conditional distribution $p(\mathbf{f}_\odot|\mathbf{y}_t,\mathbf{X}_t,\mathbf{X}_\odot)$, e.g. the distribution of $\mathbf{f}_\odot$ for observed values of $\mathbf{y}_t,\mathbf{X}_t$ and $\mathbf{X}_\odot$, is itself a multivariate Gaussian distribution:
\begin{equation}
   \mathbf{f}_\odot \mid \mathbf{y}_t, \mathbf{X}_t, \mathbf{X}_\odot \sim \mathcal{N}\left(\mathbf{K}_{t\odot}^{\top} \mathbf{K}_{tt} \mathbf{y}_t, \mathbf{K}_{\odot\odot}-\mathbf{K}_{t\odot}^{\top} \mathbf{K}_{tt}^{-1} \mathbf{K}_{t\odot}\right)
\end{equation}
where $K\left(\mathbf{X}_{t}, \mathbf{X}_{t}\right)=\mathbf{K}_{tt}$, $K\left(\mathbf{X}_{\odot}, \mathbf{X}_{t}\right)=\mathbf{K}_{\odot t}$, $K\left(\mathbf{X}_{\odot}, \mathbf{X}_{\odot}\right)=\mathbf{K}_{\odot\odot}$. The conditional mean $m(\mathbf{f}_\odot)=\mathbf{K}_{t\odot}^{\top} \mathbf{K}_{tt} \mathbf{y}_t$ is used for predictions, with the conditional covariance $\text{cov}(\mathbf{f}_\odot)=\mathbf{K}_{\odot\odot}-\mathbf{K}_{t\odot}^{\top} \mathbf{K}_{tt}^{-1} \mathbf{K}_{t\odot}$ providing an estimate of the uncertainty. The hyper parameters $l,\sigma_f^2$ that show up in the RBF kernel (eq. \eqref{RBFkernel}) can be optimized with respect to the training data, maximizing the log-likelihood (with constants omitted)
\begin{equation}
    \log p(\mathbf{y}_t \mid \mathbf{X}_t)=-\frac{1}{2} \mathbf{y}_t^{\top}\mathbf{K}_{tt}^{-1} \mathbf{y}_t-\frac{1}{2} \log \left|\mathbf{K}_{tt}\right|
\end{equation}
where it is implicit that $\mathbf{K}_{tt}$ is a function of $l$ and $\sigma_f$. The optimal parameters $l^*,\sigma_f^*$ are hence given as 
\begin{equation}
    l^*,\sigma_f^*=\arg\max_{l,\sigma_f} \left(\log p(\mathbf{y}_t \mid \mathbf{X}_t)\right)
\end{equation}
\subsubsection{Truncated sum approximation}
An alternative that does not use machine learning, but standard CC machinery, is to solve a new set of amplitude equations:
\begin{align}
    \label{eq:proj_CC_p}\bra{\Phi_0}\hat{T}^\dagger(\mathbf{x}_m)\bar H(\mathbf{c})\ket{\Phi_0}&=0 \quad \text{for all $m$}\\
    \bra{\Phi_0}\bar H(\mathbf{c})\ket{\Phi_0}&=E^*\label{eq:proj_CC_E}.
\end{align}
Eq. \eqref{eq:proj_CC_p} is motivated as follows. The CC amplitude equations \eqref{eq:CCamplitudes} can be rewritten as 
\begin{equation}
     \bra{\Phi_0}\left(\frac{\partial \hat{T}}{\partial t_\mu}\right)^\dagger\bar H\ket{\Phi_0}=0 \text{ for  $\mu\in S\cup D\cup\dots$}.
\end{equation}
Similarly, eq. \eqref{eq:proj_CC_p} corresponds to
\begin{equation}
     \bra{\Phi_0}\left(\frac{\partial \tilde{T}}{\partial c_m}\right)^\dagger\bar H\ket{\Phi_0}=0 \text{ for all $m$}.
\end{equation}Thus, projection is carried out on a set of linearly independent states that correspond to the partial derivative of the cluster operator with respect to all of its free parameters. 
This is a set of $L$ equations (compared to $O(N^4)$ in CCSD) for $L$ parameters to be solved in a similar matter as the original amplitude equations \eqref{eq:CCamplitudes}. At first glance, there is no direct numerical advantage, as one sums over all individual projection errors $\bra{\mu}\bar H(\mathbf{c}) \ket{\Phi_0}$ for $\mu\in{S\cup D\cup \dots}$:
\begin{equation}\label{eq:AMP-CCEVC-E}
    \text{e}(c_m)=\bra{\Phi_0}\hat{T}^\dagger(\mathbf{x}_m)\bar H\ket{\Phi_0}=\sum_{\mu}t_\mu(\mathbf{x}_m) \bra{\mu}\bar{H}\ket{\Phi_0},
\end{equation}
where we simply wrote $\bar H$ for $\bar H(\mathbf{c})$. However, it is not unreasonable to assume that both the error vector and the Jacobian are well approximated by considering only a small subset of the excitations, that is, assuming that 
\begin{equation}\label{eq:AMP-CCEVC-tildeE}
\begin{split}
         \text{e}(c_m)&=\sum_{\text{all }\mu {\in \{S\cup D \cup \dots\}} } t_\mu(\mathbf{x}_m) \bra{\mu}\bar{H}\ket{\Phi_0}\\
         &\approx\tilde{ \text{e}}(c_m)=\sum_{\text{some }\mu {\in \{S\cup D \cup \dots\}}} t_\mu(\mathbf{x}_m) \bra{\mu}\bar{H}\ket{\Phi_0}
\end{split}
\end{equation}
where $\tilde{ \text{e}}(c_m)$ is an approximation to the projection error. For CCSD, the AMP-CCEVC projection error reads
\begin{equation}\label{eq:FullError}
     \text{e}(c_m)\!=\!\sum_{ia} t_i^a(\mathbf{x}_m)\! \bra{\Phi_i^a}\bar{H}\ket{\Phi_0}+\frac{1}{4}\!\sum_{ijab} t_{ij}^{ab}(\mathbf{x}_m)\! \bra{\Phi_{ij}^{ab}}\bar{H}\ket{\Phi_0}\!.
\end{equation}
The most computationally expensive equations in CCSD theory in a direct product decomposition \cite{stanton1991direct} is the calculation of the $\mathcal{W}_{abef}$ intermediate and its contribution to the projection error $\bra{\Phi_{ij}^{ab}}\bar{H}\ket{\Phi_0}$, which both scale as $O(M_v^4N^2)$ :
\begin{align}
    \mathcal{W}_{abef} &=\frac{1}{4} \sum_{mn}\tau^{ab}_{mn}\braket{mn|\!|ef}+\dots \\\label{eq:WABEF}
    \bra{\Phi_{ij}^{ab}}\bar{H}\ket{\Phi_0}&= \frac{1}{2}\sum_{ef} \tau_{ij}^{ef}\mathcal{W}_{abef}+\dots
\end{align}
where we only wrote the terms scaling as $O(M_v^4N^2)$, and where
\begin{align}
    \tau_{ij}^{ab}=t_{ij}^{ab}+t_i^at_j^b-t_j^bt_i^a.
\end{align}
If we only sum over a subset of virtual indices in eq. \ref{eq:FullError}, there is no need to calculate $\bra{\Phi_{ij}^{ab}}\bar{H}\ket{\Phi_0}$  for every $a,b$, and the whole $\mathcal{W}_{abef}$ tensor is not required. Thus when only using a fraction $p\leq1$ of virtual orbitals, the number of floating point operations is, for large $M_v$, reduced by a constant factor of $\frac{1}{p^2}$ when calculating $\mathcal{W}_{abef}$ and $\bra{\Phi_{ij}^{ab}}\bar{H}\ket{\Phi_0}$. Similar considerations apply to other intermediates when not including all virtual and/or occupied orbitals. {A key observation using this method is that, by eq. \eqref{eq:ClusterOperator}, \emph{all} cluster amplitudes are updated when employing this method, hence, no excitations are excluded from the cluster operator. Only the calculation of the projection error (eq. \ref{eq:AMP-CCEVC-tildeE}) is truncated by not summing over all orbitals in all sums that enter the projection error. Thereby, the update scheme for the $c_m$ parameters is approximated, and the resulting parameters are not optimal.} It should be noted that the smallest reasonable choice of $p$ depends on the basis set employed, as well as the choice of orbitals. For larger basis sets, smaller values for $p$ are still viable, as there still will be many virtual orbitals included in the calculation. Similarly, for occupied orbitals, $p$ should be chosen in such a way that important occupied orbitals are not excluded. The truncated sum approach will yield an approximate cluster operator $\tilde{T}(\mathbf{x}_\odot)$ at reduced cost. {To exemplify the speedup using the truncated sum approximation, considering Hydrogen Fluoride in the cc-pVTZ with $p=0.1$, one calculates the projection errors and the intermediates using only 3 virtual orbitals, with the total number of virtual orbitals being 39. The $\mathcal{W}_{abef}$ tensor is thus only calculated for $a,b<3$, and other intermediate tensors in the direct sum decomposition are also only needed for $a,b<3$.} In order to decide which orbitals to include, one can use the \emph{average relative importance} of an orbital, which we define for occupied and virtual orbitals, respectively, as
\begin{align}
        \Theta_i&=\sum_{m=1}^L\sum_{abj}\abs{t_{ij}^{ab}(\mathbf{x}_m)}^2 \label{eq:Truncation_occ}\\ \Theta_a&=\sum_{m=1}^L\sum_{bij}\abs{t_{ij}^{ab}(\mathbf{x}_m)}^2. \label{eq:Truncation_virt}
\end{align}
A large value $\Theta_a$ means that double excitations into orbital $a$ are important, and correspondingly, a large value $\Theta_i$ means that double excitations from orbital $i$ are important. For a given $p$, one can then use only those virtual orbitals with the largest $\Theta_a$ and those occupied orbitals with large $\Theta_i$. It should be noted that, just like in regular CC theory, DIIS \cite{diis,scuseria_diis} can be used to speed up solving for the approximate cluster operator. \\\\
We will now justify why valid results are to be expected when this approximation is used. Assume the AMP-CCEVC approach to be exact, e.g. assume there is a parameter $\mathbf{c}$ that exactly solves the CC amplitude equations
\begin{align}
    &\bra{\mu}e^{-\tilde{T}(\mathbf{x}_\odot)}\hat{H}e^{\tilde{T}(\mathbf{x}_\odot)}\ket{\Phi_0}=0 \quad\text{for all }\mu \in S\cup D \cup \dots \label{eq:AMP_perf} \\
   & \tilde{T}(\mathbf{x}_\odot)=\sum_{m=1}^Lc_m\hat{T}(\mathbf{x}_m).
\end{align}
Then the amplitude equations are fulfilled for any arbitrary linear combinations of states, e.g.
\begin{equation}
    \sum_{\text{some } \mu}t_\mu\bra{\mu}e^{-\tilde{T}(\mathbf{x}_\odot)}\hat{H}e^{\tilde{T}(\mathbf{x}_\odot)}\ket{\Phi_0}=0
\end{equation}
independently of the value and the choice of $t_\mu$. Thus, both eqs. \eqref{eq:AMP-CCEVC-E} and \eqref{eq:AMP-CCEVC-tildeE} will be equal to zero and thus share the same solution. If the AMP-CCEVC approach does not solve eq. \eqref{eq:AMP_perf} exactly, the truncated sum approach may have different roots than AMP-CCEVC with $p=1$. If the AMP-CCEVC approach is approximately correct, the solutions of eq. \eqref{eq:AMP-CCEVC-tildeE} will be close to those of eq. \eqref{eq:AMP-CCEVC-E}. When the AMP-CCEVC amplitudes are not a good approximation to the CC amplitudes, we expect the truncated sum solution to be qualitatively different from both the untruncated AMP-CCEVC and the CC solution.
\subsection{Molecular orbitals at different geometries}
In second quantization, the Hamiltonian with nuclei placed at positions $\vec{{R}}$ reads
\begin{equation}
\begin{split}\label{eq:Hamiltonian}
        \hat{H}(\vec{R})
    &=h_{\text{nuc}}(\vec{R}) + \sum^M_{p q} h_{p q}(\vec{R}) a_{p}^{\dagger}(\vec{R}) a_{q}(\vec{R})\\&+\frac{1}{4} \sum^M_{p q r s}\braket{pq|\!|rs}\!\!(\vec{R}) a_{p}^{\dagger}(\vec{R}) a_{q}^{\dagger}(\vec{R}) a_{s}(\vec{R}) a_{r}(\vec{R}).
\end{split}
\end{equation}
The one-body integrals, the two-body integrals and the creation and annihilation operators depend on the nuclear positions. Still, following the discussion in Ref. \cite{helgaker_second-quantization_1984}, one might consider the creation and annihilation operators to be geometry-independent, using an occupation number (ON) vector representation, which is geometry-independent. The HF state, for example, can be represented as $\ket{1_1\dots 1_N0_{N+1}\dots0_M}$ in ON vector representation, independently of geometry. This geometry independent representation can be used for all $2^M$ basis states of the Fock space. As long as the MOs are orthogonal at every geometry, one can thus consider the Fock spaces at different geometries as the same Fock space. Doing this, creation and annihilation operators can be considered as constant entities that do not depend on geometry, thereby removing the geometry dependence from the creation and annihilation operators. {It should be noted that in order for the matrix representation in Slater-determinant basis of eq. \eqref{eq:Hamiltonian} to be continuous as a function of nuclear positions $\Vec{R}$, such that the eigenvectors can be continuous, we require $h_{p q}(\vec{R})$ and $\braket{pq|\!|rs}\!\!(\vec{R})$ to be continuous, which corresponds to requiring continuous MOs.}
\subsection{Procrustes orbitals}
The (converged) cluster amplitudes $\mathbf{t}(\vec{R})$ at different nuclear geometries $\vec{R}$ depend not only on $\vec{R}$ and the choice of the basis set, but also a choice of of both the occupied orbitals in the reference Slater determinant $\ket{\Phi_0}$ as well as the virtual orbitals. AMP-CCEVC requires the cluster operators at all geometries to refer to excitations from the same set of occupied orbitals into the same set of virtual orbitals at all geometries, e.g. for both sample and target cluster operators.
This requires a particular choice of molecular orbitals. 
\subsubsection{Motivation}\label{sec:Procrustes_motivation}
\begin{figure}[h!]
    \centering
    \includegraphics[width=0.5\textwidth]{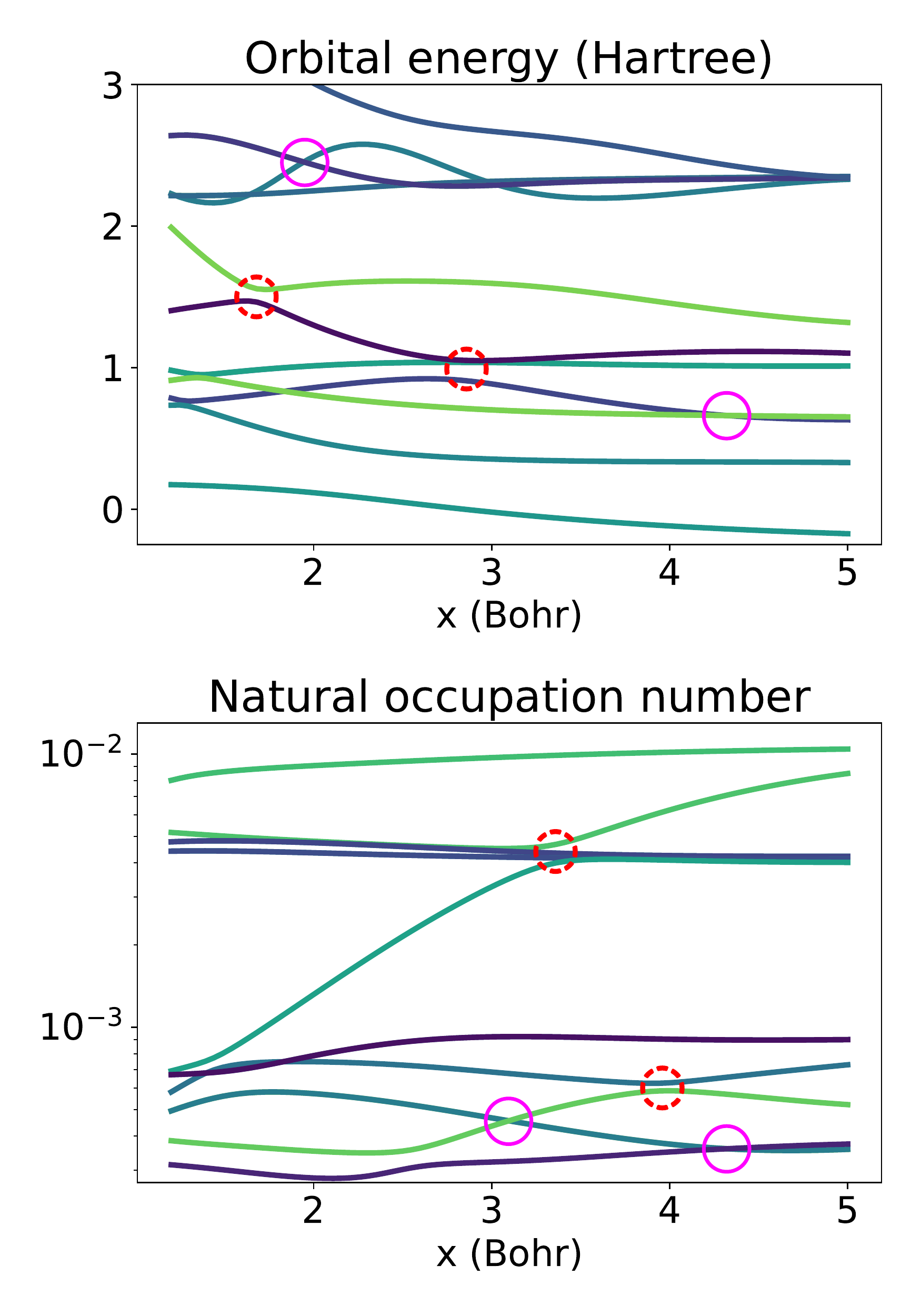}
    \caption[Crossings and avoided crossings]{Trajectory of some orbital energies (top) and natural occupation numbers (bottom) as function of internuclear distance $x$ for hydrogen fluoride in the cc-pVTZ basis set. Some avoided crossings are marked in red (with dashed lines), while proper crossings are marked in pink.}
    \label{fig:avoided_crossings}
\end{figure}

A re-indexing of MOs leads to problems. For example, when two canonical orbitals $\ket{\phi_i},\ket{\phi_j}$ switch index $i \leftrightarrow j$ as the nuclei are moved, which might happen in standard quantum chemistry programs when they cross in energy for orbitals ordered by molecular energy, $\ket{\phi_i(\Vec{R})}$ is no longer smooth (or even continuous) as a function of $\vec{R}$, and neither is the orbital energy \cite{Kato}. Those crossings occur for both canonical and natural orbitals, which is exemplified by the hydrogen fluoride molecule in the cc-pVTZ basis set in  figure \ref{fig:avoided_crossings}. In this case, $\mathbf{t}(\vec{R})$ will not be continuous.
While it is possible to keep track of those crossings, it requires a relatively high resolution of the PES at Hartree-Fock level. The situation is made worse by avoided crossings between MOs obtained as solution of a matrix diagonalization, such as canonical orbitals or natural orbitals (see fig. \ref{fig:avoided_crossings}). Assume there is an avoided crossing between $\ket{\phi_i(\Vec{R})}$ and $\ket{\phi_j(\Vec{R})}$. Even though $\ket{\phi_i(\Vec{R})}$ is smooth, $\ket{\phi_i(\Vec{R}-\Delta \vec R)}$ resembles $\ket{\phi_j(\Vec{R}+\Delta \vec R)}$ instead of $\ket{\phi_i(\Vec{R}+\Delta \vec R)}$, where $\Vec{R}-\Delta \vec R$ is one one side of the avoided crossing and $\Vec{R}+\Delta \vec R$ on the other \cite{Avoided_crossings}. This problem cannot be solved by reordering MOs and requires a choice of a new set of orbitals without (avoided) crossings, obtained through orbital rotations between occupied orbitals as well as virtual orbitals, without mixing. {A possible alternative to finding a new type of MOs would be to define a transformation that maps a sample cluster operator from one geometry to another, taking into account how the MOs have changed. However, we have no knowledge of how such a transformation might look like - the best we can do is find a set of suitable MOs.}
\subsubsection{Definition}
{
Let $\mathbf{C}(\vec{R})^{(o)}$ and $\mathbf{C}(\vec{R})^{(v)}$ describe the coefficient matrices for the occupied/virtual canonical MOs at geometry $\vec{R}$. The \emph{Procrustes orbitals} with respect to reference geometry $\vec{R}'$ are represented by the MO coefficient matrices as
\begin{equation}
    \begin{split}
    \mathbf{C}_{\vec{R}'}(\vec{R})^{(o)}=\mathbf{C}(\vec{R})^{(o)}\mathbf{Q}^{(o)}\\
    \mathbf{C}_{\vec{R}'}(\vec{R})^{(v)}=\mathbf{C}(\vec{R})^{(v)}\mathbf{Q}^{(v)}\\
    \end{split}
\end{equation}
where $\mathbf{Q}^{(o)/(v)}$ are chosen in such a way that they make the occupied/virtual orbitals at geometry $\Vec{R}$ as close to the orbitals at geometry $\Vec{R}'$ as possible. In order to measure closeness between orbitals, we can define the matrix $\mathbf{W}^{(o)}(\vec{R})$ with orthogonal columns
\begin{equation}\label{eq:W_matrix_o}
    \mathbf{W}^{(o)}(\vec{R})=\mathbf{S}(\vec{R})^{\frac{1}{2}}\mathbf{C}^{(o)}(\vec{R})
\end{equation}
where $\mathbf{S}(\vec{R})$ is the atomic orbital (AO) overlap matrix at geometry $\Vec{R}$. $\mathbf{W}(\vec{R})$ tells how the occupied MOs expressed using the "intrinsic" MO coefficient matrix $\mathbf{S}(\vec{R})^{-\frac{1}{2}}$ (which resemble the AOs the most) are to be rotated to obtain canonical orbitals. Similarly, we can define $\mathbf{W}^{(v)}$
\begin{equation}\label{eq:W_matrix_v}
    \mathbf{W}^{(v)}(\vec{R})=\mathbf{S}(\vec{R})^{\frac{1}{2}}\mathbf{C}^{(v)}(\vec{R})
\end{equation}
In order to make the MOs at geometry $\Vec{R}$ as close to the orbitals at geometry $\Vec{R}'$ as possible, we can find unitary matrices ${\mathbf{Q}}^{(o)}/{\mathbf{Q}}^{(v)}$ that make the MOs as similar as possible in Frobenius norm $\left\|\cdot\right\|_F$:
\begin{equation}\label{eq:Procrustes_alt}
\begin{split}
{\mathbf{Q}}^{(o)}=\arg \min _{\mathbf\Omega}\left\|\mathbf{W}(\vec R)^{(o)}\mathbf{\Omega}-\mathbf{W}(\vec R')^{(o)}\right\|_{F} \hspace{0.5em} \text {s.t.} \hspace{0.5em} \mathbf\Omega^{\dagger} \mathbf\Omega=\mathbf{1}\\
{\mathbf{Q}}^{(v)}=\arg \min _{\mathbf\Omega}\left\|\mathbf{W}(\vec R)^{(v)}\mathbf{\Omega}-\mathbf{W}(\vec R')^{(v)}\right\|_{F} \hspace{0.5em} \text {s.t.} \hspace{0.5em} \mathbf\Omega^{\dagger} \mathbf\Omega=\mathbf{1}
\end{split}
\end{equation}
Thus, the Procrustes orbitals become
\begin{equation}
    \begin{split}
    \mathbf{C}_{\vec{R}'}(\vec{R})^{(o)}=\mathbf{S}(\vec{R})^{-\frac{1}{2}}\mathbf{W}^{(o)}(\vec{R}){\mathbf Q}^{(o)}=\mathbf{C}(\vec{R})^{(o)}\mathbf{Q}^{(o)}\\
    \mathbf{C}_{\vec{R}'}(\vec{R})^{(v)}=\mathbf{S}(\vec{R})^{-\frac{1}{2}}\mathbf{W}^{(v)}(\vec{R}){\mathbf Q}^{(v)}=\mathbf{C}(\vec{R})^{(v)}\mathbf{Q}^{(v)}.
    \end{split}
\end{equation}
The optimization problem in eq. \eqref{eq:Procrustes_alt} is known as the \emph{orthogonal Procrustes problem} and has an analytical, unique solution \cite{Procrustes}.
 \begin{equation}
\begin{split}
         \mathbf{M}^{(o)}&=\qty(\mathbf{W}(\vec R)^{(o)}) ^\dagger \mathbf{W}(\vec R')^{(o)}\\
         &=\qty(\mathbf{C}^{(o)}(\vec{R}))^\dagger\qty(\mathbf{S}(\vec{R})^{\frac{1}{2}})^\dagger \mathbf{S}(\vec{R'})^{\frac{1}{2}}\mathbf{C}^{(o)}(\vec{R}')\\
         &=\mathbf{U}^{(o)}\mathbf{ \Sigma}^{(o)}\left(\mathbf{V}^{(o)}\right)^\dagger
\end{split}
\end{equation}
 and using the Singular Value Decomposition (SVD) \cite{Banerjee2014}, the solution is given by 
\begin{equation}
    \mathbf{Q}^{(o)}=\mathbf{U}^{(o)}\left(\mathbf{V}^{(o)}\right)^\dagger
\end{equation}
and similarly for $\mathbf{Q}^{(v)}$. {Observe that even when singular values in $\mathbf{ \Sigma}^{(o)}$/$\mathbf{ \Sigma}^{(v)}$ should coincide, thereby making the matrices $\mathbf{U}^{(o)}/\mathbf{U}^{(v)}$ and $\mathbf{V}^{(o)}/\mathbf{V}^{(v)}$ ambiguous, the product $\mathbf{Q}^{(o)}/\mathbf{Q}^{(v)}$ remains unique, so this causes no issues.} Brillouin's theorem still applies to Procrustes orbitals.}
\subsubsection{Considerations regarding AMP-CCEVC}
For AMP-CCEVC, we will use Procrustes orbitals at all geometries $\vec{R}$ to ensure that cluster operators at different geometries refer to the same orbitals. We have observed that the choice of the reference geometry $\vec{R}'$ is essentially arbitrary{, as we have found that the Procrustes orbitals at geometry $\Vec{R}$ are close to the Procrustes orbitals at geometry $\Vec{R}'$ when $\Vec{R}$ and $\Vec{R}'$ are close, independently of what the reference geometry is. This is the case when the Procrustes orbitals are continuous, which they are unless the reference determinant changes abruptly.} It should be noted that many CCSD-algorithms rely on an approximate, diagonal, constant Jacobian $\mathbf{J}$ with elements
\begin{equation}\label{eq:Jacobian}
    J_{\mu\nu}(\mathbf{t})=\epsilon_\mu\delta_{\mu\nu}
\end{equation}
where $\epsilon_\mu=\epsilon_{aibj}=\epsilon_{a}+\epsilon_{b}-\epsilon_{i}-\epsilon_{j}$ and $\epsilon_{ai}=\epsilon_{a}-\epsilon_{i}$ is the difference between the orbital energies \cite{bible}. This approximation is only reasonable when using  canonical orbitals or orbitals similar to canonical orbitals, thus, the (approximate) amplitudes w.r.t. Procrustes orbitals need to be transformed back to canonical orbitals if they are to serve as an initial guess, which is an easy undertaking scaling only as $O(N^5)$. {The average relative importance of an orbital (eqs. \eqref{eq:Truncation_occ} and  \eqref{eq:Truncation_virt}) is calculated using the sample amplitudes obtained using Procrustes orbitals in order to decide which orbitals are included when calculating the approximate error \eqref{eq:AMP-CCEVC-tildeE} and thereby the update scheme for the $c_m$ parameters. That is, in the truncated sum approximation, only those orbitals are used in the update scheme that have the highest average importance, considering all sample cluster operators obtained using Procrustes orbitals.}
\subsection{Input parameters for GP for molecular geometries}
The value of the RBF kernel at two different molecular geometries (eq. \eqref{RBFkernel}) depends on the specification of $d(\vec{R},\vec{R}')=\norm{h(\vec{R})-h(\vec{R}')}$ (eq. \eqref{eq:distance}). A natural choice would be $h(\vec{R})=\vec{R}$. For nonlinear molecules, however, this representation becomes ambiguous, does not take into account translational and rotational symmetry, and does not encode the behaviour of the molecular orbitals. In machine learning for quantum chemistry, descriptors for molecular geometry such as Coulomb matrices \cite{CoulombMatrix} and Bags of Bonds \cite{BoB} (which are identical to Coulomb matrices when considering the same molecule) are designed to predict energies and expectation values across very different molecules. We have found that these predictors do not work well, as they do not encode information about the orbitals, which is what is needed. As the aim is to predict particular linear combinations of cluster operators, an orbital-dependent representation is needed.
{A reasonable choice to measure similarity is to measure intrinsic properties of the MOs using the matrices introduced in eqs. \eqref{eq:W_matrix_o} and \eqref{eq:W_matrix_v}. Defining the unitary matrix $\mathbf{W}(\vec R)=[\mathbf{W}^{(o)}(\vec R)\quad \mathbf{W}^{(v)}(\vec R)]$, it measures how the MOs are constructed from the "intrinsic" MOs and, being unitary, has a Frobenius norm that is bounded. Thus, we use as a distance measure
\begin{equation}\label{eq:Distance_U}
    d(\vec{R},\vec{R}')=
    \left\|\mathbf{W}(\vec{R})-\mathbf{W}(\vec{R}')\right\|_{F}.
\end{equation}
}
For the RBF kernel (eq. \eqref{RBFkernel}), we found improved results by setting a lower bound of $1.3$ for the hyper parameter $l$ to avoid overfitting.
\subsection{Choice of sample geometries}
We consider here two different choices for the sample geometries: The first is to simply use a uniformly spaced grid. This is what we will do to obtain the sample cluster operators in the truncated sum approach.\\
However, we also consider using the information about the uncertainty in the predictions from the Gaussian model. That is, we devise an algorithm on how to decide the $(k+1)$th sample geometry based on the $k$ previously determined Gaussian models. If one has the Hartree-Fock state available at all geometries of interest, or if there is an estimate of the kernel matrix $K(\mathbf{X}_\odot,\mathbf{X}_t)$ for all geometries of interest {(which, for example, can be obtained using a HF--calculation in minimal basis or interpolation) }, one can calculate the insecurity in the estimate $\sigma_i(\vec{R})$ for $i=1,\dots,k$ at all geometries of interest. One can then pick the geometry  $\hat{R}$ with the largest cumulative insecurity, e.g. with largest value for 
\begin{equation}
    \hat{R}=\arg\max_{\vec{R}}\left(\sum_{i=1}^{k}\sigma^2_i(\vec{R})\right)
\end{equation}
as the geometry with the least knowledge, where the covariance is ignored. One can then obtain the cluster operator at that geometry, either by obtaining the full cluster operator or using AMP-CCEVC, and add it to the set of sample cluster operators. Then, new Gaussian processes need to be fit to the data. As reasonable starting geometries, we recommend using the $2^d$ corner points of the grid, where $d$ is the dimension of the geometric perturbation considered. 

\section{Results}\label{sec:Results}
{All results presented here used DIIS to obtain the converged cluster operator. For calculations using the cluster operator from the previous geometry, we used Procrustes orbitals with respect to that previous geometry (e.g. with no fixed reference), as other orbital choices might be insufficient by the discussion of sec. \ref{sec:Procrustes_motivation}, and because the previous geometry is a natural choice for the reference geometry. We used Procrustes orbitals, not alternative Procrustes orbitals, for all calculations.}
\subsection{Number of iterations for CC calculations}
\begin{figure}[t!]
    \centering
    \includegraphics[width=0.5\textwidth]{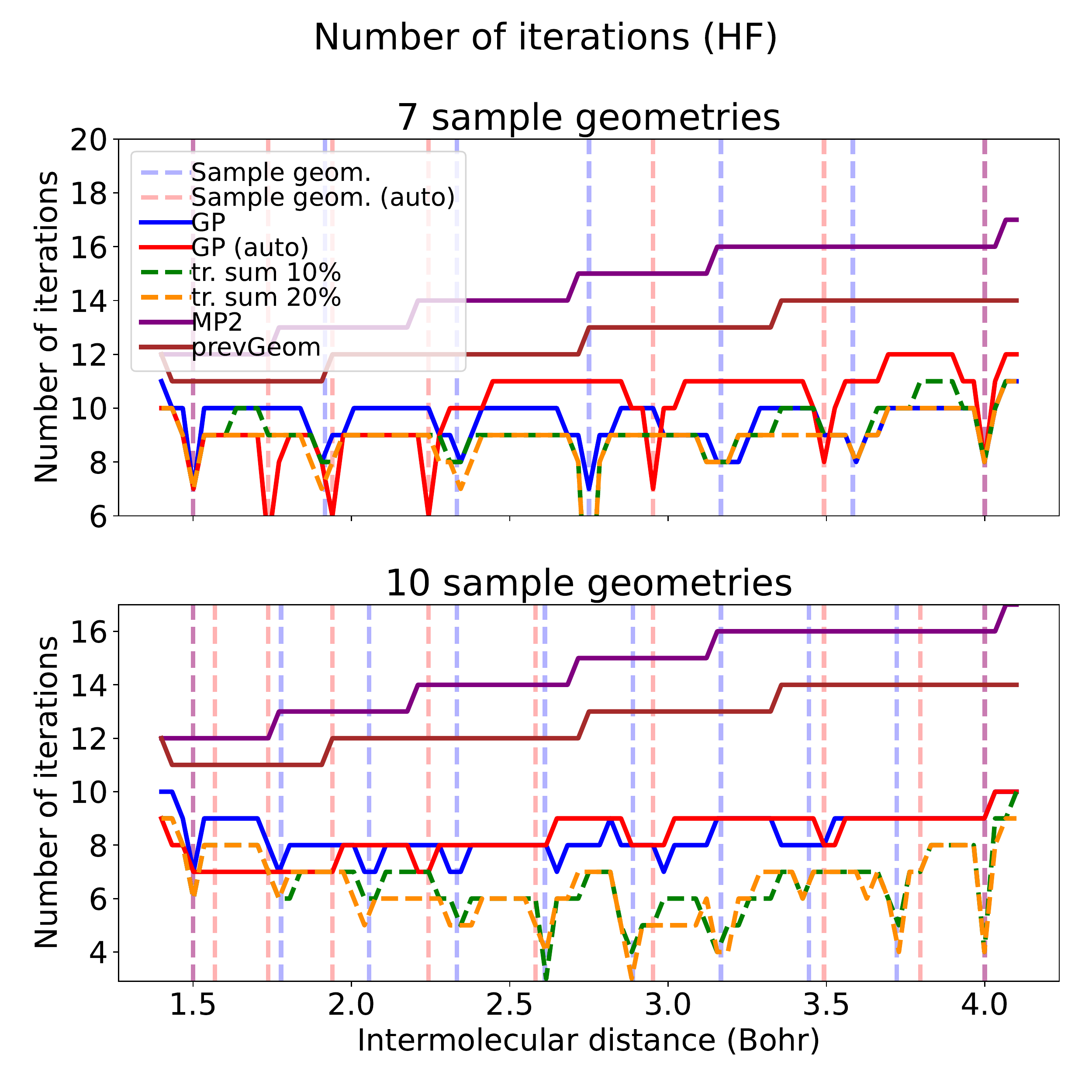}
    \caption{The number of iterations necessary to converge the maximal projection error to $10^{-8}$ using an MP2 start guess, the previous geometry ($\Delta R =0.03375$) (with Procrustes orbitals), the truncated sum approximation ($p=0.1,0.2$) (here called "AMP") and Gaussian processes using either a grid or automatic selection for the geometries, using 7 and 10 sample geometries for hydrogen fluoride in the cc-pVTZ basis set. The equilibrium geometry $R_0=1.75$ (Bohr) was chosen as the reference geometry. Violet dashed lines indicate that the sample geometries coincide.}
    \label{fig:HF7iter}
\end{figure}
Figure \ref{fig:HF7iter} shows the number of iterations for the hydrogen fluoride molecule in the cc-pVTZ basis set using a standard MP2 guess, the previous geometry ($\Delta R =0.03375$) (with Procrustes orbitals), the truncated sum approximation ($p=0.1$,$p=0.2$) and Gaussian processes. Uniformely spaced sample geometries are used. We also considered the performance with sample geometries based on the Gaussian process. We used 7 and and 10 sample geometries, respectively. The number of iterations compared to the MP2 guess is reduced in all cases, and both the machine-learning ansatz and the truncated sum ansatz give superior convergence, especially in the out-of-equilibrium region, where the MP2 guess becomes worse. There is also improved convergence compared to using the previous geometry. As expected, increasing the number of sample geometries reduces the number of iterations further, with, on average, well below 10 iterations using 10 sample geometries with all methods considered. The dips close to the sample geometries can be explained by the fact that the start guess already is very close, or equal, to the exact cluster operator. For Gaussian processes, one might expect convergence after one iteration. However, for numerical stability, a small number ${\epsilon=10^{-10}}$ was added to the diagonal of the covariance matrix in order to make it invertible, thereby giving slightly different values than the exact sample values. This also explains why the energy in the next section is not necessarily exact at the sample geometries. The truncated sum approach gives results that beat the machine learning approach, and using $p=0.2$ leads to fewer iterations compared to $p=0.1$. Both are not surprising, given that the truncated sum approach actively solves a set of equations to obtain the cluster operator, while the machine learning approach simply estimates the value of the underlying complicated functions, and using larger $p$ gives improved approximations to the sum. Finally, we observe that the automatic sampling procedure {using 7 sample geometries} leads to a higher number of iterations, which is because sampling is denser close to the equilibrium region, which is related to larger changes in the unitary rotation matrix $\mathbf{W}(\vec{R})=\mathbf{S}^{1/2}(\vec{R})\mathbf{C}(\vec{R})$. {This difference, however, disappears as more sampling geometries are added.} 
\begin{figure}[h!]
    \centering
    \includegraphics[width=0.5\textwidth]{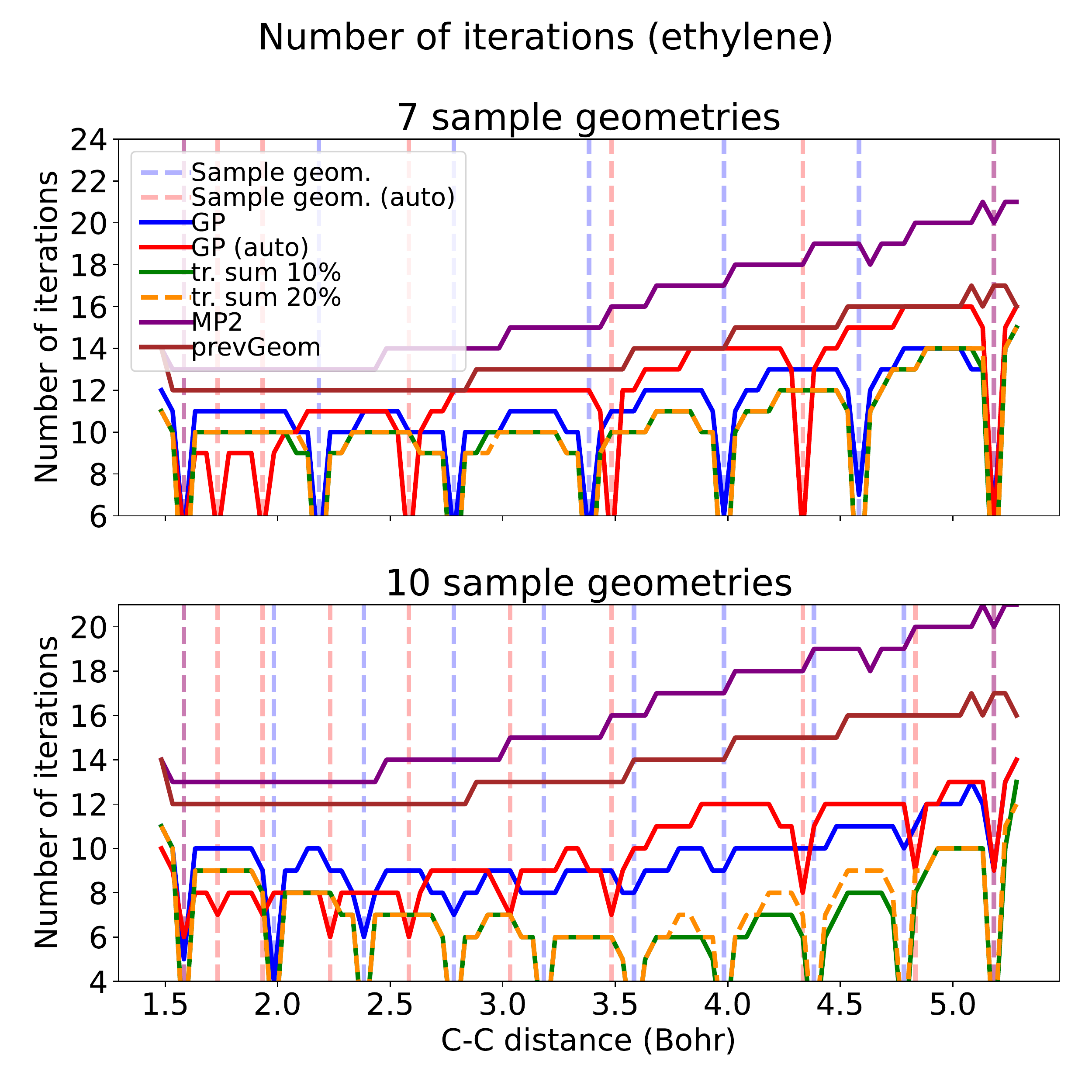}
    \caption{The number of iterations necessary to converge the maximal projection error to $10^{-8}$ using an MP2 start guess, the previous geometry ($\Delta R =0.05$) (with Procrustes orbitals), the truncated sum approximation ($p=0.1$) (here called "AMP") and Gaussian processes using either a grid or automatic selection for the geometries, using 7 and 10 sample geometries for the linear bond break in ethylene in the cc-pVDZ basis set. The equilibrium geometry was chosen as Procrustes reference geometry. Violet dashed lines indicate that the sample geometries coincide.}
    \label{fig:ethene_iter}
\end{figure}

As a second test system, we considered the \emph{linear} dissociation of the ethylene C=C bond (e.g. stretching the bond, with the bond length $x$ as variable of interest). The number of iterations is shown in figure \ref{fig:ethene_iter}. The observations are essentially the same as in the case of hydrogen fluoride, but it shows in addition that AMP-CCEVC can be applied also to describe bond breaking situations that are more complex than breaking a single, two-atom bond, with an even more drastic improvement in the number of iterations as three extra sample geometries are added. {We also observe that the truncated sum approach works slightly better with $p=0.1$. While this is a surprising result, we believe that this is a lucky coincidence, with the approximate cluster operator being coincidentally closer to the exact one. The difference in the number of iterations is only one step, so the difference between the cluster operators is small.} It should be noted that using a maximal projection error of $10^{-8}$ leads to an extremely accurate cluster operator. For energy calculations, this type of accuracy is not required. By decreasing this number, the absolute improvement in the number of iterations remains approximately the same, thereby leading to a larger relative improvement in the number of iterations.
\subsection{Approximate potential energy surfaces}
\begin{figure}[h!]
    \centering
    \includegraphics[width=0.5\textwidth]{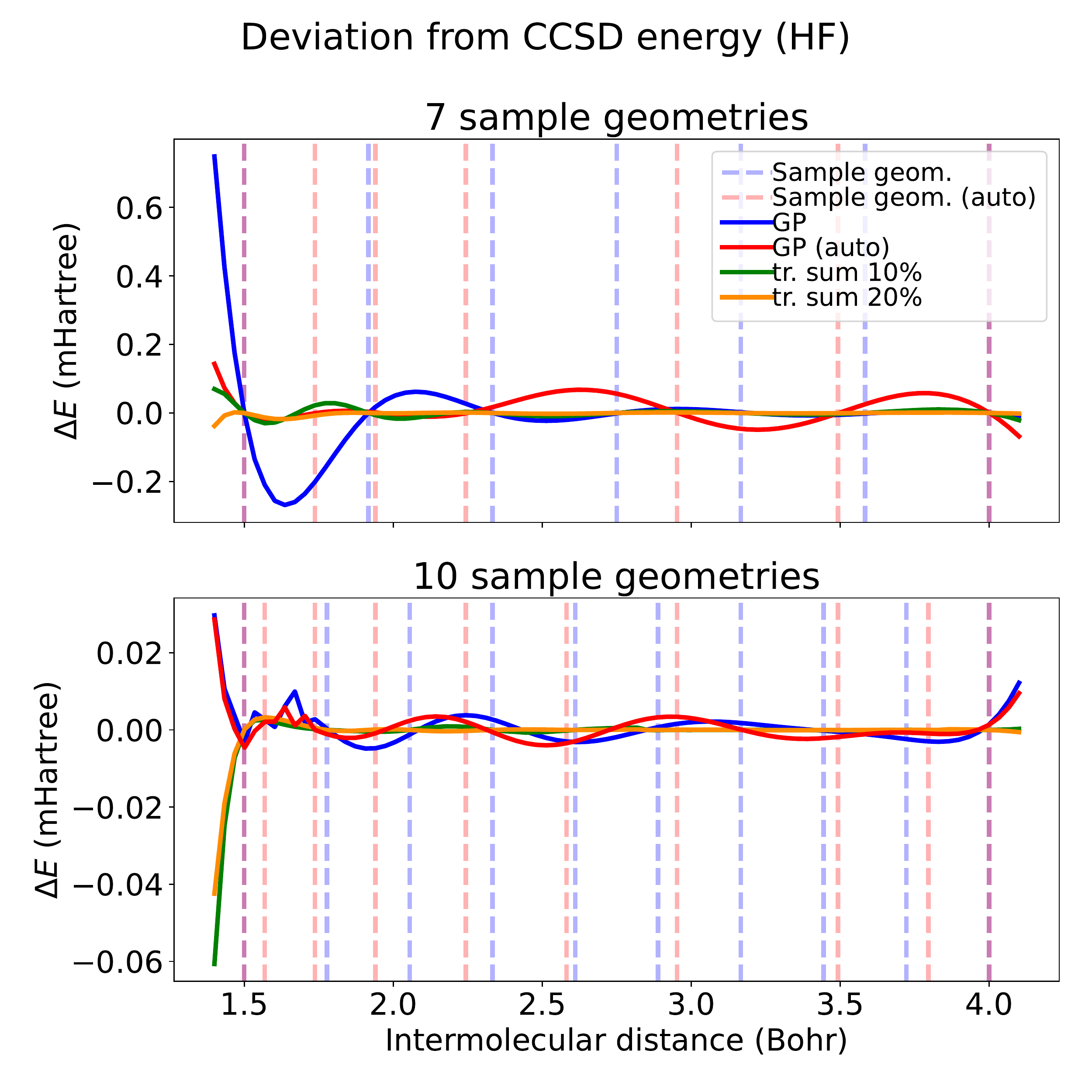}
    \caption{Error compared to the exact CCSD energy using the truncated sum approximation ($p=0.1,0.2$) and Gaussian processes with evenly spaced sample points or automatic sample point selection with 7 and 10 sample geometries for hydrogen fluoride in the cc-pVTZ basis set. Violet dashed lines indicate that the sample geometries coincide.}\label{fig:HFenergy}
\end{figure}
Instead of using AMP-CCEVC to obtain a good starting guess for the true cluster operator, one might also use the AMP-CCEVC cluster operator to obtain an approximate potential energy. We illustrate this for hydrogen fluoride using 7 and 10 sample geometries in figure \ref{fig:HFenergy}.
\begin{figure}[h!]
    \centering
    \includegraphics[width=0.5\textwidth]{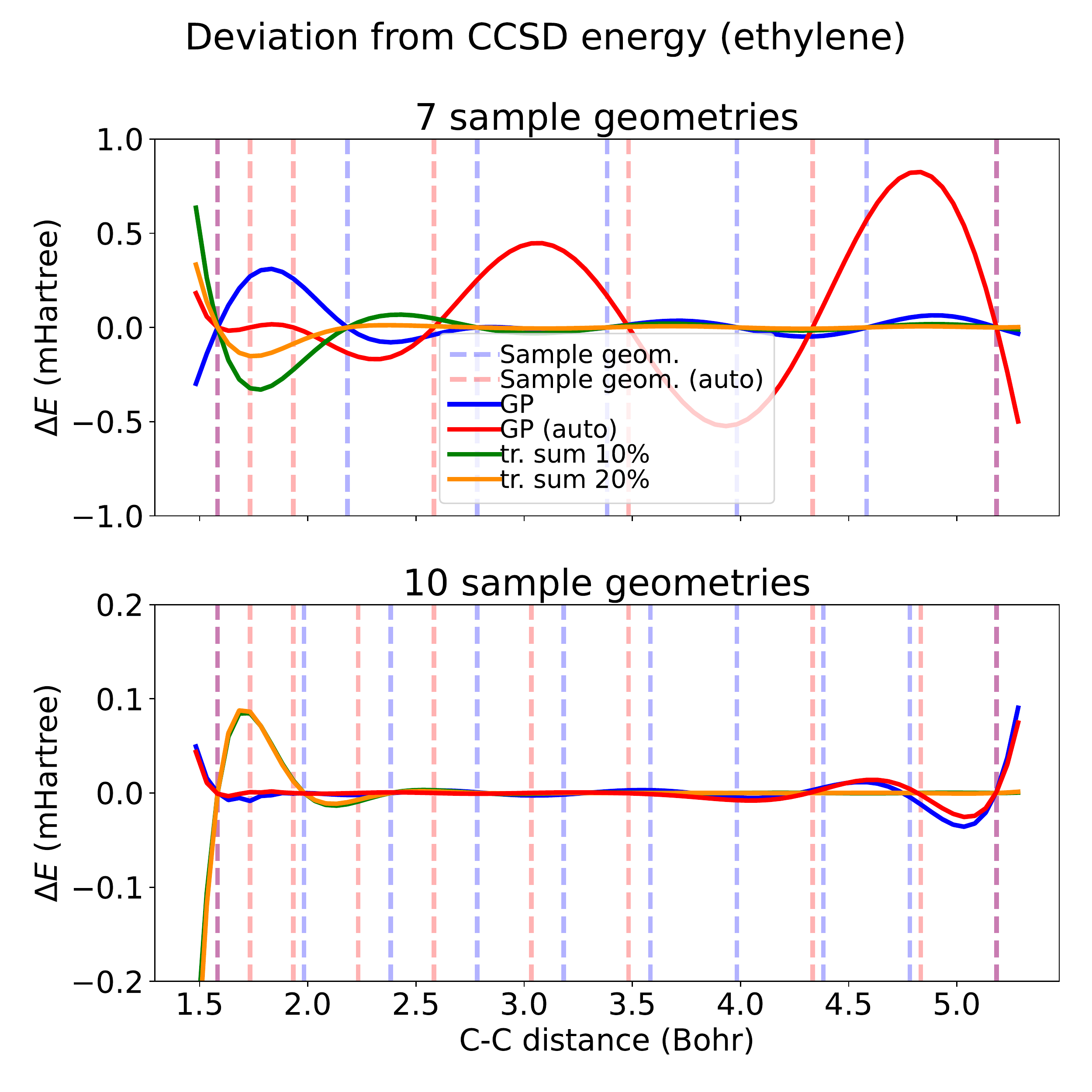}
    \caption{Error compared to the exact CCSD energy using the truncated sum approximation ($p=0.1,0.2$) and Gaussian processes with evenly spaced sample points or automatic sample point selection with 7 and 10 sample geometries for ethylene in the cc-pVDZ basis set.The equilibrium geometry was chosen as Procrustes reference geometry. Violet dashed lines indicate that the sample geometries coincide.}\label{fig:ethylene_energy}
\end{figure}
Again, we observe that the truncated sum approach works better than using Gaussian processes, however, the results lie well within chemical accuracy for all methods considered. There is an improvement by an order of magnitude by increasing the number of sampling points from 7 to 10.  We observe the largest deviations from the CCSD energy close to the equilibrium geometry (1.4 to 2 Bohr) with particularly large deviations in the extrapolation region for the $p=0.1$ sum approximation and the GP approach when using a uniform grid, with clear improvements when increasing that number to $10$. As expected, using $p=0.2$ gives better results than using $p=0.1$. In the automatic selection of the sample geometries, going from 7 to 10 sample geometries, sample geometries were added very close to where the energy deviated maximally from the CCSD energy, showing that the automatic selection algorithm is efficient. {In table \ref{tab:HF_errors}, we show the agreement with the coupled cluster correlation energy, e.g. the average value for \begin{equation*}
    \%_{\text{corr.}}=100\%\cdot \left(1-\left|\frac{{\Tilde{E}_{CC}}}{{E_{CC}}-E_{HF}}\right|\right),
\end{equation*} where $\Tilde{E}_{CC}$ stands for the approximative energy, where the average is taken over all geometries considered. This shows that all methods manage to recover more than $99.9\%$ of the CCSD correlation energy, with the best methods even recovering $99.999\%$.
} 
\begin{table}[h!]
\centering
\caption[Agreement with correlation energy]{{Average agreement $\%_{\text{corr.}}$ with CCSD correlation energy for the hydrogen fluoride molecule, sampled at 81 evenly spaced geometries between $R=1.4$ Bohr and $R=4.1$ Bohr as shown in figure \ref{fig:HFenergy}. The number in parentheses refers the number of sample geometries.}}{\begin{tabular}{lll}
\hline
Method    & $\%_{\text{corr.}}$ (7) & $\%_{\text{corr.}}$ (10) \\ \hline
tr. sum (10\%)    & 99.9969  &  99.9995                \\ 
tr. sum (20\%)    &  99.9991 &          99.9996           \\ 
GP        & 99.9819 &     99.9991                           \\ 
GP (auto) & 99.9903 &     99.9992                     \\      
\end{tabular}}
\label{tab:HF_errors}
\end{table}
\begin{table}[h!]
\centering
\caption[Agreement with correlation energy]{{Average agreement $\%_{\text{corr.}}$ with CCSD correlation energy for the ethylene molecule, sampled at 77 evenly spaced geometries in $[-1+R_0,2.8+R_0]$ (Bohr) for the distance between the carbon atoms, where $R_0$ is the equilibrium geometry, as shown in figure \ref{fig:ethylene_energy}. The number in parentheses refers the number of sample geometries.}}
{\begin{tabular}{lll}
\hline
Method    & $\%_{\text{corr.}}$ (7) & $\%_{\text{corr.}}$ (10) \\ \hline
tr. sum (10\%)    & 99.9811  &  99.9961                \\ 
tr. sum (20\%)    &  99.9924 &          99.9959           \\ 
GP        & 99.9814 &     99.9981                           \\ 
GP (auto) & 99.9235 &     99.9985                     \\      
\end{tabular}}
\label{tab:etylene_errors}
\end{table}

The corresponding energy deviations for the dissociation of ethylene is shown in figure \ref{fig:ethylene_energy}, {with the average agreement with the CCSD correlation energy shown in table \ref{tab:etylene_errors}}.
There is a clear improvement in the ML approach, going from 7 to 10 sample geometries. Using 10 sample geometries, both are very accurate, with deviations from the CCSD energy by less than {0.1} mHartree in the interpolation region, and still small for extrapolation, {compared to errors up to almost 1 mHartree using 7 sample geometries}. Observe in particular the improvement between 3.5 and 4.5 Bohr for the automatic-sampling approach, even though no additional geometries were added there (but at the other two maxima). {We also observe that the automatic sampling approach performs worse using 7 sample geometries, as it tends to prefer sampling close to the equilibrium geometry. However, with 10 sample geometries, it performs better than the grid. Again, all methods recover more than $99.9\%$ of the CCSD correlation energy.}

\begin{figure*}[ht!]
    \centering
    \includegraphics[width=0.8\textwidth]{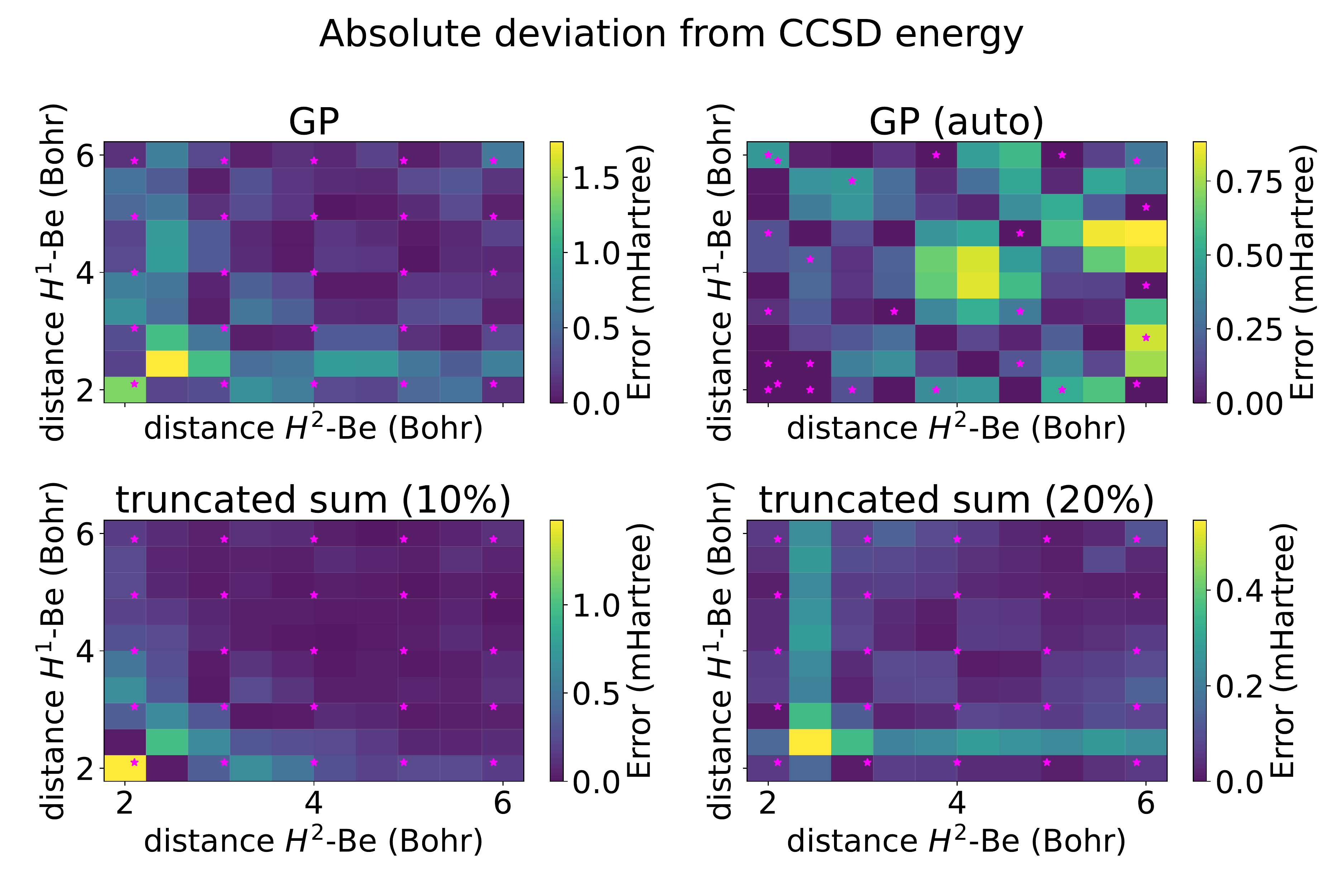}
    \caption{Error compared to the exact CCSD energy using the truncated sum approximation ($p=0.1,0.2$) and Gaussian processes with evenly spaced sample points or automatic sample point selection with 25 sample geometries for the asymmetric stretch of \ch{BeH2} in the cc-pVTZ basis set. Sample geometries are marked as pink stars. The {reference} geometry was chosen as $d(Be,H_1)=d(Be,H_2)=2$ Bohr.}\label{fig:BeH2_energy}
\end{figure*}
\subsection{Two-dimensional case}

We consider again the number of iterations and the deviation from the CCSD energy, this time for a two-dimensional model. We look at the asymmetric stretch of \ch{BeH2}, where the molecule remains linear, but the bond between $\ch{Be}$ and each $\ch{H}$ is stretched independently. Figure \ref{fig:BeH2_energy} depicts the deviation from the exact CCSD energy, using 25 sample geometries. Here, we relaxed the convergence condition of the maximal projection error to $10^{-6}$.

We observe again that the automatic selection of the geometries for the Gaussian process prefers the equilibrium region. We observe that the largest absolute error for the {grid} is essentially chemical accuracy, while it is { better for the other methods, with very good results for the GP with automatic selection of grid points, and the truncated sum approximation with 
 $p=0.2$.} We see that using 25 sample geometries is {already sufficient to} obtain chemical accuracy. {This is especially surprising as the average grid distance is 1 Bohr, which is considerably more than in the previous examples.} Increasing the number of sample geometries would give improved energies. Nevertheless, we have found that even with only 25 sample geometries, all of the methods considered reduced the number of iterations (compared to the MP2-guess) consistently. Specifically, table \ref{tab:Niter_BeH2} presents the average number of iterations {as well as the average agreement with the CCSD correlation energy} using the different {methods}, { with the automatic selection procedure and the truncated sum approximations reducing the number of average iterations by around 4, and recovering at least $99.7\%$ of the correlation energy.} 
\begin{table}
\centering
\caption[Number of iterations for \ch{BeH2}]{Average number of CCSD iterations {$n_{\text{it.}}$} to reduce maximal projection error to $10^{-6}$ for the asymmetric stretch of \ch{BeH2} using different methods, {as well as average agreement $\%_{\text{corr.}}$ with the CCSD correlation energy. The values} are averaged over the same grid as the one shown in figure \ref{fig:BeH2_energy}, e.g. 100 points uniformly distributed on the grid $[2,6]^2$, and the same sample geometries as in figure \ref{fig:BeH2_energy} are used.}
\begin{tabular}{lll}
\hline
Method    & $n_{\text{it.}}$ & $\%_{\text{corr.}}$ \\ \hline
MP2       & 13.2                         &        -          \\ 
tr. sum (10\%)    & 9.3                  &        99.8479           \\ 
tr. sum (20\%)    & 9.1                  &        99.9098           \\ 
GP        & 10.2                          &        99.6701          \\ 
GP (auto) & 9.4                          &        99.7705        \\      
\end{tabular}
\label{tab:Niter_BeH2}
\end{table}
\section{Discussion}\label{sec:Discussion}
We have seen that AMP-CCEVC can be used to get an approximately correct cluster operator. Using the truncated-sum approach, fewer sample geometries are necessary to get close to the correct cluster operator than when using Gaussian processes, especially for larger values of the percentage of included orbitals $p$, at the expense that the truncated sum approach formally still scales as $O(N^6)$. For the machine-learning approach with Gaussian processes, we have seen that an average spacing between points of $\sim 0.5$ Bohr is sufficient to reproduce the potential energy surface to chemical accuracy, and just slightly increasing this number can give energies that are accurate to within 0.1 mHartree to the CCSD energy. Furthermore, we have found that one can use the uncertainty estimates from Gaussian processes to obtain sample geometries. When using a grid to obtain sample geometries, AMP-CCEVC cannot be used to escape the curse of dimensionality, as the number of grid points increases exponentially with the number of geometric dimensions. It is however possible that using the automatic selection algorithm might give sub-exponential scaling. 
\subsection{Possible use}
AMP-CCEVC can be used to obtain an initial guess of the cluster operator which converges faster to the cluster operator than using a "naive" MP2 parameter guess. A significant advantage of the method is that it is straightforward to increase the accuracy by adding additional sampling geometries. The automatic selection algorithm in the machine-learning approach gives a rule which geometry to add. When adding points in the ML-approach, one has the choice of whether to use a "full" cluster operator, or whether to use the truncated sum approach (which we have not considered here). In addition, the method can be used in such a way that it provides approximate CCSD energies, which, when enough sample cluster operators are used, can provide energy estimates which differ from the CCSD energy by less than chemical accuracy. As this only requires a HF-reference and Procrustes orbitals and an evaluation of the CCSD energy, the machine learning ansatz gives approximate CCSD energies with $O(N^5)$ scaling. 
\subsection{Restrictions}
\begin{figure}[hbt!]
        \centering
        \includegraphics[width=0.3\textwidth]{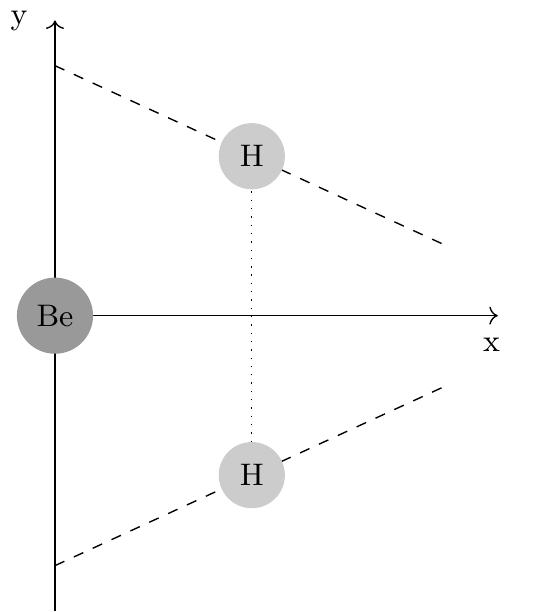}
        \caption[\ch{BeH2} insertion reaction - illustration]{The \ch{BeH2} insertion reaction. Beryllium is placed in the origin, the position of the hydrogen atoms as functions of a parameter $x$ is given by $R_x=x$, $R_y=\pm \left(2.54 - 0.46x\right)$ (in Bohr).}
        \label{fig:BeH2_drawing}
    \end{figure}
CC theory is a single-reference method. As such, the energy is discontinuous when the reference determinant is changed discontinuously as function of the nuclear geometry, and so are the amplitudes $\mathbf{t}$. Thus, AMP-CCEVC is not applicable whenever the reference determinant changes discontinuously along a path, for the same reason that CC theory fails. Prime examples are the chromium dimer and the \ch{BeH2} insertion reaction \cite{BeH2_insertion,Bodenstein}, where \ch{Be} is inserted into \ch{H2} (see fig. \ref{fig:BeH2_drawing}). The Hartree--Fock state with the lowest energy changes discontinuously at $x\approx2.85$ Bohr (the exact value depending on the basis set), which also leads to a discontinuous change in the Procrustes orbitals and the cluster operator as function of $x$. This discontinuous change makes the cluster operator change discontinuously. Thus, it is no longer reasonable to assume that the cluster operator for  $x>2.85$ can be written as a linear combination of cluster operators for  $x<2.85$. 
\subsection{Future research}
As the evaluation of the CCSD energy scales as $O(N^4)$ independently of the truncation level of the cluster operator, the methods discussed here are not restricted to CCSD, but can be applied to CCSD(T), CCSDT and other truncation levels. AMP-CCEVC has applications outside of quantum chemistry and can be used in fields of physics that make use of CC theory. It can also be used for other types of perturbations, such as external electromagnetic fields. More generally, the method has likely a good transferability to NISQ algorithms on quantum computers, where obtaining the best parameters $\vec \theta$ for a parameterized unitary operator $\hat U (\vec \theta)$ acting on a reference state $\ket {\Phi_0}$ is not straightforward and no natural choice of starting guesses exists. Similarly to CC theory, MP2 amplitudes can be used for unitary coupled cluster theory \cite{Romero_2018}, with the same advantages and drawbacks as in regular CC theory. The machine learning approach can be further improved by the choice of a different, possibly tailored kernel $k(\vec{R},\vec{R}')$. Furthermore, we believe that the method can be applied to (approximate) geometry optimization. For the truncated sum approach approach, we believe that further improvements are possible by (approximately) minimizing the sum of squared projection errors  
\begin{equation}
     \mathbf{c}=\argmin_{\mathbf{c}^*} \sum_{\mu} \norm{\bra{\mu}\bar H(\mathbf{c}^*)\ket{\Phi_0}}^2.
\end{equation}
In addition to obtaining an approximate cluster operator $\hat{T}$, the methods considered here can in a straightforward way also be extended to the $\lambda$-equations in CC theory, making it possible to calculate approximate response properties and expectation values at reduced cost. Furthermore using sparse grids \cite{SparseGrids} for multidimensional problems might give a polynomial scaling in the number of sample geometries, avoiding the curse of dimensionality. A further research goal is to adapt AMP-CCEVCC to multi-reference coupled cluster methods \cite{Piecuch1993_JCP,Ivanov2009_PCCP}. Given that the used reference state is continuous, AMP-CCEVC can be applicable to these type of problems.  
{Finally, further developing the method to be compatible with localized orbitals, such as pair natural orbitals \cite{PNOs} or local natural orbitals \cite{LNOs,LNOs2}, might lead to large speed ups. Such a development can also make it possible to use the methods described here in each physical domain of the localized space individually, which also would allow for a transferability of localized cluster operators between molecules of different size and composition.}
\section{Conclusion}\label{sec:Conclusion}
For calculating the cluster operator, we have shown that cluster operators obtained at previously calculated geometries can be used to reduce the number of calculations necessary to converge the cluster operator. We introduced Procrustes orbitals, obtained by transforming the canonical orbitals, which are a set of molecular orbitals that changes as little as possible when atoms in a molecule are moved. This makes it possible to reuse calculations from a previous geometry also in situations where using canonical or natural orbitals would be complicated due to (avoided) crossings. By using AMP-CCEVC, where the target cluster operator is written as a linear combination of sample cluster operators, we found that the number of iterations can be significantly diminished by reusing the cluster operators from more than one previous geometry, either by explicitly solving a reduced number of amplitude equations, or by learning the best linear combination of the sample cluster operators. For the latter part, we found that Gaussian processes work well, as they can predict the expansion coefficients based on the similarity between molecular orbitals, and because they allow for zero training error without overfitting. Furthermore, we found that both approaches can be used to obtain potential energy surfaces that are chemically exact within CCSD accuracy. In particular, we found that as few as 7 sample cluster operators, selected automatically, are sufficient to describe a double bond dissociation reaction going over several Ångstrøm, with the main cost being to obtain those 7 sample cluster operators, with a further improvement by more than an order of magnitude by increasing that number to 10. As it is not the energy, an expectation value, but the cluster operator itself which is being interpolated, one can also use this method to calculate arbitrary expectation values.
{
\section*{Data availability}
A repository with the code, containing an implementation of AMP-CCEVC, which we used to create all data and figures except for figures 1 and 7, is available on Github \cite{ProcrustesCode}. Our code builds on and requires the locally developed and openly available HyQD Python package~\cite{QuantumSystems,CCcode} for CC calculations, and on the PySCF Python package \cite{pySCF1,pySCF2} for quantum chemical integral evaluation and Hartree--Fock calculations.}
\section*{Author Declaration}
The authors have no conflicts to disclose.
\section*{Acknowledgements}
The work was supported by the Research Council of Norway through its Centre of Excellence funding program, Project No. 262695.
\bibliography{main}

\end{document}